# Evaluation of the "disorder temperature" and "free volume" formalisms via simulations of shear banding in amorphous solids

Yunfeng Shi, Michael B. Katz, Hui Li and Michael L. Falk

Materials Science and Engineering

University of Michigan

Ann Arbor, MI 48109-2136 USA

*March 20, 2007*

ABSTRACT

Molecular dynamics simulations of shear band development over 1000% strain in simple shear are used to test whether the local plastic strain rate is proportional to $\exp(-1/\chi)$, where $\chi$ is a dimensionless quantity related to the "disorder temperature" or "free volume" that characterizes the structural state of the glass. Scaling is observed under the assumption that $\chi$ is linearly related to the local potential energy per atom. We calculate the potential energy per atom corresponding to absolute zero disorder temperature, and the energy needed to create a shear transformation zone.

Being able to correctly model the physics of plastic deformation is critically important for understanding the mechanical failure modes of materials. Research into the dynamics of plasticity in metals has traditionally considered the dynamics of crystalline dislocations [1]. However a significant number of complications arise in the analysis of heavily dislocated and polycrystalline solids, and the systematic development of a statistical mechanics of deformation based on dislocation plasticity remains a grand challenge. Less attention has been paid to the plastic responses of metallic glasses [2-8]. Yet insight into the statistical mechanics of plastic flow in such amorphous solids may aid in the construction of theories of plasticity in general.

In previous studies of the system simulated here we observed that samples prepared at sufficiently slow quench rates undergo an instability resulting in two coexisting regions: a shear band that contributes nearly all the plastic flow, and a jammed region that undergoes negligible plastic deformation[9, 10]. The bifurcation of the mechanical response exhibited in simulation appears to be closely related to the development of shear bands in metallic glasses. Being that these shear bands represent the primary failure mode in that material system they have been extensively studied experimentally [11-15]. Here we drive the system at constant strain rate in a "simple shear" geometry to achieve very large strains. This loading geometry results in a uniform stress across the sample, and allows us to average across parallel layers to collect adequate statistics for verifying theoretical predictions.

It is important to note that during plastic deformation significant amounts of heat are generated. However, recent experimental investigations indicate that the origin of shear bands in metallic glasses is not a thermo-mechanical softening of the material, as in high rate of deformation in steels, but rather it arises due to structural softening [16, 17]. After shear band slip has initiated, however, heating may play a significant role in late stages of slip. The purpose of this investigation is to consider the processes that control shear band initiation, and in order to minimize the thermal effects we purposely simulate a system in which cooling is applied at a high rate throughout the sample.

There presently exist a number of theoretical predictions for the mechanical response of metallic glasses [ 2, 5, 7, 18, 19], making it difficult to perform a complete comparative study. Instead we choose to consider one theoretical assumption present in a wide variety of theories[ 2, 18, 19], but which, as far as we are aware, has never been directly validated in experiment or simulation. These theories assume that the structural states of the glass, which have fallen out of equilibrium with the bath, obey Poisson statistics. The earliest instance of this assumption of which we are aware is the theory of Cohen and Turnbull[20] used by Spaepen to construct his theory of metallic glass deformation[2]. Theories that follow in this vein[ 19, 21] assume that the distribution of "free volume" controls the flow rate. Recently Langer [18] and Lemaitre [22] have questioned whether a quasi-thermodynamic interpretation is preferable. Simulation results have indeed illustrated that zero temperature systems under shear may behave in a way typically associated with thermally equilibrated systems, i.e. they may be well described by an "effective temperature" [23, 24]. It is of considerable theoretical interest to determine which of these assumptions hold.

In order to analyze our simulation results we will assume that, since the simulations are performed at very low temperature, the rate of thermal relaxation of the structural state is negligible outside the shear band. Also, we will assume that any internal degrees of freedom, other than a single scalar parameter that governs the distribution of structural states, equilibrate quickly by comparison. For example in the context of the shear transformation zone (STZ) theory an evolving distribution of defects known as STZs naturally result in phenomena such as jamming and hysteresis [5]. Here we assume that the number and orientations of the STZs maintain their steady state values. We will also assume convection is negligible.

In this limit the free volume theory that Heggen, Spaepen and Feuerbacher applied to model the behavior of $Pd_{41}Ni_{10}Cu_{29}P_{20}$ [8], the theory of Lemaitre and Carlson [21] and the disorder temperature theory of Langer [18] reduce to nearly the same functional form. These consist of two equations, one that relates the local plastic strain rate, $\dot{\varepsilon}_{pl}$, to the shear stress, $s$, and the structural state, $\chi$, through an equation of the form

$$\dot{\varepsilon}_{pl} = e^{-1/\chi} f(s), \qquad (1)$$

and a second equation that provides the evolution of the structural state,

$$c_0 \dot{\chi} = 2s\dot{\varepsilon}_{pl}(\chi_\infty - \chi) - k(\chi, T). \qquad (2)$$

In the above equations $f(s)$ is a monotonic function of stress; the parameter $c_0$ is a quantity with units of energy density; the dimensionless parameter $\chi_\infty$ is the upper limiting value of $\chi$, and the function $k$ is the rate of thermal relaxation, discussed below. Although these theories lead to similar equations, the interpretation of $\chi$ differs. In the disorder temperature formalism this parameter is defined as $\chi \equiv k_B T_d / E_Z$, where $k_B$ is the Boltzmann factor; $T_d$ is the disorder temperature that characterizes energy distribution of configurations in the glass, and $E_Z$ is a typical energy required to create an STZ. In the free volume model $\chi \equiv v_f / \gamma v^*$ is known as the reduced free volume, where $v_f$ is the free volume, $v^*$ is a critical free volume and $\gamma$ is a parameter of order unity. This difference is not, in principle, irreconcilable since the free volume could be related in important ways to the disorder temperature.

Despite their similarities the theories cited above are distinct in notable ways. Eq. (2) implies that in flowing regions $\chi$ converges to a limiting value $\chi_\infty$ in the absence of thermal relaxation. This feature is present only in the analysis by Langer. The second difference arises in the functional form of $k$, the rate of thermal relaxation, that depends on both $\chi$ and the temperature. Spaepen and Heggen, *et al.* assume due to binary annihilation $k \propto -e^{-1/\chi}\left(e^{-1/\chi} - e^{-E_Z/kT}\right)$. Langer proposes that $k \propto -e^{-\beta/\chi}\left(\chi - \frac{kT}{E_Z}\right)$, which is similar to the form employed by Lemaitre, *et al.* Since we are operating at very low temperature and our interest is confined to relaxations occurring inside the shear band, both can be approximated as

$$k(\chi,T) \approx \kappa \exp(-\beta/\chi) \qquad (3)$$

where $\kappa$ is a kinetic constant, and $\beta$ is a parameter. The third distinguishing feature is the form of $f(s)$, which exhibits a dynamical transition from jamming to flow at finite stress only in the STZ theory. However, the form of $f(s)$ has no consequence in this analysis.

We have performed molecular dynamics simulation in a two-dimensional model of a binary alloy that was originally developed in the context of quasi-crystals [25]. We have simulated this system in numerous other investigations of the plastic deformation of glasses, and refer the reader to these sources for the details of the atomic interactions and the composition of the system [5, 9, 10]. All energies are expressed in terms of $\varepsilon_{SL}$, the energy of the bond between an atom of the two species, denoted $S$ for small and $L$ for large. Lengths are expressed in terms of a reference length scale, $\sigma_{SL}$, the distance at which the $SL$ interaction energy is zero. The mode coupling temperature, a good upper bound for the glass transition temperature, has been calculated to be $T_{MCT} \approx 0.325\, \varepsilon_{SL}/k$. Temperatures will be expressed in terms of $T_{MCT}$. All the particles have the same mass, $m_0$. The reference time scale will therefore be $t_0 = \sigma_{SL}\sqrt{m_0/\varepsilon_{SL}}$.

The systems used here span a square box 286.26 $\sigma_{SL}$ on a side and consist of 80,000 atoms with 35,776 $L$ atoms and 44,224 $S$ atoms. The initial conditions were created starting from supercooled liquids equilibrated at 1.08 $T_{MCT}$. The temperature of the liquid was then reduced to 9.2% of $T_{MCT}$ isochorically. We begin the mechanical tests from nine independent glass configurations created by lowering the temperature over one of three different quenching times, which we will denote Quench I (100,000 $t_0$), Quench II (50,000 $t_0$) or Quench III (10,000 $t_0$). In this way we are able to consider sample-to-sample variations. The average potential energies per atom are -2.507, -2.497 and -2.477 $\varepsilon_{SL}$ for Quench I, II and III, respectively.

Simple shear is induced by incrementally deforming the simulation cell. To avoid the artifacts such a transformation causes at very large strain, we integrate the SLLOD equations of motion [26] and apply Lees-Edwards boundary conditions [27]. Full periodicity is maintained along all boundaries. In keeping with the simple shear geometry all shear stresses and shear strain rates will refer to the magnitude of the shear in the direction of the applied loading, i.e. $s_{xy}$ and $\dot{\varepsilon}_{xy} \equiv \partial^2 u_x / (2\partial y \partial t)$.

During shearing, the temperature is maintained at 9.2% $T_{MCT}$ by coupling to a Nose-Hoover thermostat [28]. This can be interpreted as analogous to cooling a thin film sample from the exposed surface. Due to the high rate of cooling temperatures remain well below the glass transition temperature. Three strain rates are employed: 0.00002, 0.0001 and 0.0005 $t_0^{-1}$. For the lowest shear rates, the temperature in the shear band rises to about 11% $T_{MCT}$, only slightly higher than the surrounding material. Even for the fastest shear rates, the temperatures never reach even 1/3 of $T_{MCT}$. Thus the material in the band remains well below the glass transition temperature. Careful analysis indicates that 98-99% of the mechanical work done on the system is dissipated as heat. This is in agreement with the experimental estimate of Heggen, et. al. [8]. The maximum shear strain is 1000%. Fig. 1 shows the stress strain curves for one sample prepared at each of Quench I, II and III and tested at a strain rate of 0.00002 $t_0^{-1}$.

In order to investigate the effects of strain rate and sample preparation, we calculated the distribution of shear strain at 50% total applied strain using the method detailed in [5] as shown in Fig. 2. High strain rate promotes homogenous deformation via multiple shear bands, consistent with our observations in previous uniaxial tensile tests [10]. The deformation in Fig. 2 is quite similar for Quench I, II and III due to a narrower range of quenching rates than our previous parametric study.

Once initiated, shear bands typically lead to fracture, but in simple shear the shear band can evolve to extremely high strains. In Fig. 3a we have quantified the evolution of the strain rate profile in a shear band by measuring displacements of atoms in 50 horizontal slices at strain intervals of 100%. Fig. 3b demonstrates that we can also detect the band by examining the local potential energy per atom.

In the disorder temperature model $T_d$ is a temperature in some thermodynamically meaningful sense. Consequently it should be possible to relate $\chi$ to the local potential energy. Here we assume that $\chi$ is a linear function of the potential energy per atom.

$$C_1 \chi = PE - PE_0 \qquad (4)$$

where $PE$ is the potential energy; $PE_0$ is the potential energy per atom for ideally jammed regions, and $C_1$ is a quantity with units of energy per atom that is equal to $m_0 E_Z / k$ times a specific heat that relates the internal energy to the disorder temperature.

When the system is loaded in simple shear, the material inside and outside the band are held at the same average shear stress by construction due to mechanical equilibrium. Therefore if we divide the strain rate at a given $y$ position by the strain rate at the center of the shear band, $\dot{\varepsilon}_b$, where the value of $\chi$ is equal to $\chi_b$, Eq. (1) predicts

$$\frac{\dot{\varepsilon}_{pl}(y)}{\dot{\varepsilon}_b} = exp\left(\frac{1}{\chi_b(s)} - \frac{1}{\chi(y)}\right). \qquad (5)$$

Note that because $f$ is only a function of the shear stress, $s$, this undetermined function drops out of the equation except in determining $\chi_b$. By solving Eqs. (1)-(3) for the steady state value of $\chi$, the value of $\chi_b$ can be expressed as a function of the strain rate in the band. Taking this expression and Eqs. (4) and (5) we find that

$$ln\left[\frac{\dot{\varepsilon}_{pl}(y)}{\dot{\varepsilon}_b}\right] - \frac{1}{\chi_\infty - r\dot{\varepsilon}_b^{-1}} = \frac{C_1}{PE_0 - PE(y)} \qquad (6)$$

Here we have expanded the value of $\chi_b$ around $\chi_\infty$ assuming high strain rates in the band; $r$ is the resulting rate constant that depends on $\kappa$, $\chi_\infty$, $\beta$ and the functional form of $f$.

This scaling relationship should hold irrespective of the stage of the deformation process or the location in the material.

Fig. 4 shows the left-hand side of Eq. (6) as determined from the local strain rates plotted versus potential energy. Each data point is calculated from the strain rate and potential energy of horizontal slices of material 5.72 $\sigma_{SL}$ thick. Slices are taken from simulations every 100% strain from 200% to 1000%. The data is averaged over 50 data points ordered by potential energy per atom. This includes data extracted from simulation results for all three initial conditions from each of the three quench times. Each pane represents the data from a different applied strain rate. Here we have estimated $\chi_\infty$ to be 0.1548 and *r* to be $3.565 \times 10^{-4} t_0^{-1}$. Data collapse is evident for data from all 9 simulations at all strains over a 3 order of magnitude range of local strain rates. The solid line in each figure represents the functional form shown on the right-hand side of Eq. (6) when we estimate **PE₀** to be -2.58 $\varepsilon_{SL}$ and **C₁** to be 1.15 $\varepsilon_{SL}$. Our ability to fit nearly all the data using Eq. (6) provides strong confirmation of the theoretical framework discussed above. The one systematic exception occurs at the highest shear rates for the highest and lowest potential energies. This data comes from the earliest strains. As illustrated in Fig. 2, multiple shear bands are evident in these cases. So this deviation should be expected.

Through our analysis of these simulations, a surprisingly consistent picture of plastic deformation in glassy materials appears to emerge. The ability of Eq. (1) to describe the deformation behavior when the state variable $\chi$ is interpreted as being linearly related to the potential energy per atom is consistent with a quasi-thermodynamic interpretation of this quantity. Edwards, Langer and Lemaitre have all argued for the existence of such a quantity, related to the entropy, that characterizes the number of paths available to the system[18, 22, 29]. This conceptual framework could form the basis of a thermodynamics of the "jammed" state of glasses and granular materials. Such an interpretation would potentially shed light on the value of $PE_0$, the degree of structural relaxation corresponding to "absolute zero" disorder temperature. This ideally jammed state could be "randomly dense packed" or crystalline. The crystalline ground state for this system has a PE of -2.60 $\varepsilon_{SL}$, very close to our $PE_0$ value of -2.58 $\varepsilon_{SL}$.

Our analysis has indicated that in the regime of importance for these simulations the potential energy per atom and the free volume are linearly related. From this one could argue that the two formalisms are interchangeable. However, it is only in the context of the thermodynamic interpretation that the value of the energy per STZ can be extracted from $\chi_\infty$. If we make the assumption that $T_d$ must be equal to $T_g \approx T_{MCT}$ in the shear band in the limit of high strain rate, then we can estimate $E_Z \approx kT_{MCT}/\chi_\infty = 2.1\varepsilon_{SL}$, about 2-4 atomic bonds in strength. This energy scale is consistent with the STZ concept as originated by Argon [3], and is crucially important for understanding the kinds of elementary processes that control deformation.

The authors would like to acknowledge support of the NSF under award DMR-0135009 and PHY99-07949. In addition we benefited from discussions with J.-L. Barrat, J.S. Langer, A. Lemaitre and P. Sollich.

Figure captions

Figure 1 Stress-strain curves from Quenches I, II and III tested at a strain rate 0.00002 $t_0^{-1}$ up to 20% strain. The inset shows strains up to 1000%.

Figure 2 Distributions of plastic deformation in a sample from Quench I (left), Quench II (center) and Quench III (right) at 50% strain. Three strain rates are shown for each: 0.00002 (top), 0.0001 (middle) and 0.0005 $t_0^{-1}$ (bottom). Black denotes strains in excess of 50% shear strain; white denotes 0% strain.

Figure 3 The strain rate (a) and potential energy per atom (b) as a function of the vertical position in the simulation cell measured at various strains for a sample from Quench I sheared at a strain rate of 0.00002 $t_0^{-1}$.

Figure 4 The scaled local strain rate as a function of potential energy for all data from Quench I, Quench II and Quench III. The panes show data from each of the three applied strain rates. The error bars are calculated by averaging over fifty data points sorted by potential energy per atom. Only data from 200% to 1000% strain are included.

Figure 1

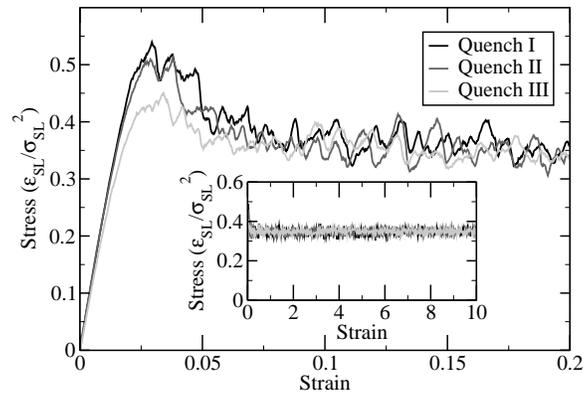

Figure 2

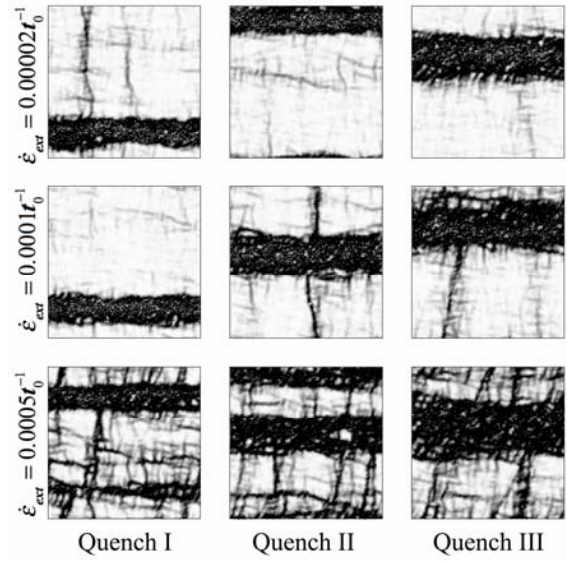

Figure 3

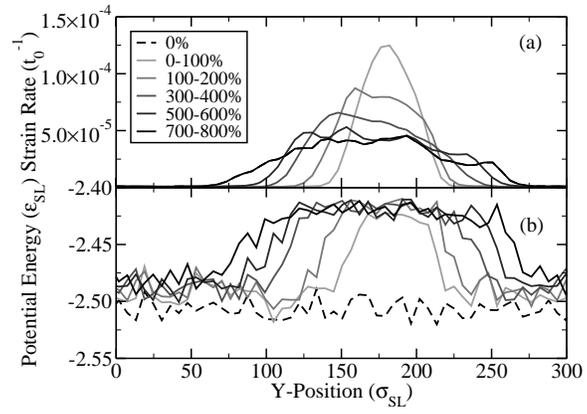

Figure 4

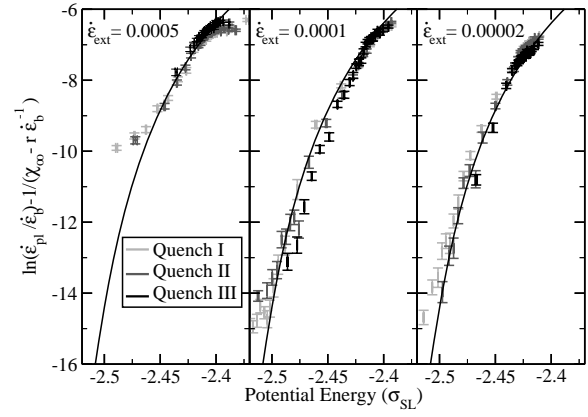